# A General Approach to Compute Phosphorescent OLED Efficiency

*Xu Zhang,[a] Denis Jacquemin,[b,c] Qian Peng,[d]\* Zhigang Shuai,[a]\* Daniel Escudero[b]\**

[a] MOE Key Laboratory of Organic OptoElectronics and Molecular Engineering, Department of Chemistry, Tsinghua University, Beijing 100084, People's Republic of China. E-mail: zgshuai@tsinghua.edu.cn;

[b] CEISAM UMR CNRS 6230, Université de Nantes, 2 rue de la Houssinière, BP 92208, 44322 Nantes Cedex 3, France. E-mail : Daniel.Escudero@univ-nantes.fr;

[c] Institut Universitaire de France, 1, rue Descartes, 75005 Paris Cedex 5, France

[d] CAS Key Laboratory of Organic Solids, Institute of Chemistry of the Chinese Academy of Sciences, Beijing 100190, People's Republic of China. E-mail: qpeng@iccas.ac.cn

**Abstract:** Phosphorescent organic-light emitting diodes (PhOLEDs) are widely used in the display industry. In PhOLEDs, cyclometalated Ir(III) complexes are the most widespread triplet emitter dopants to attain red, e.g., Ir(piq)$_3$ (piq = 1-phenylisoquinoline), and green, e.g., Ir(ppy)$_3$ (ppy = 2-phenylpyridine) emission, whilst obtaining operative deep-blue emitters is still one of the major challenges. When designing new emitters two main characteristics, besides colours, should be targeted: high photostability and large photoluminescence efficiencies. To date, these are very often optimized experimentally in a trial-and-error manner. Instead, accurate predictive tools would be highly desirable. In this contribution, we present a general approach for computing the photoluminescence lifetimes and efficiencies of Ir(III) complexes by considering all possible competing excited state deactivation processes, and importantly explicitly including the strongly temperature-dependent ones. This approach is based on state-of-the-art quantum chemical calculations with excited state decay rate formalism and the kinetic modelling, which is shown to be both efficient and reliable for a broad palette of Ir(III) complexes, i.e., from yellow/orange to deep-blue emitters.

**Introduction**

Molecular materials modeling and virtual screening approaches are becoming fundamental tools in the design of new molecules with tailored properties,[1,2] including electroluminescent molecules.[3,4] To boost the interest of high-throughput virtual screening approaches, robust computational approaches accurately predicting all key properties are first needed. PhOLEDs, the so-called second generation of OLEDs,[5] are still the most widespread OLEDs technology in display industry. Therefore, enormous industrial and academic efforts are devoted to the development of new electroluminescent molecular materials with optimal performances for PhOLEDs, and in particular to the elaboration of highly-efficient and photostable deep-blue phosphors. Recent reports showed that external quantum efficiencies (EQE) of up to 30% can be obtained using horizontally oriented heteroleptic Ir(III) complexes.[6,7,8] Due to the intricate nature of the competing excited state deactivation processes in Ir(III) complexes,[9,10] the accurate computation of the photoluminescence efficiency, which is the main factor determining the EQE values in PhOLEDs, has remained elusive up to now. In practice, the efficient triplet harvesting requires an emissive triplet excited state presenting a substantial metal-to-ligand charge transfer ($^3$MLCT) character,[11,12] in order to favor the radiative drainage to the electronic ground state over the nonradiative one. In addition, the photoluminescence lifetimes and efficiencies are temperature-dependent,[10] due to the presence of thermally activated nonradiative channels associated to the population of triplet metal-centered ($^3$MC) states. The competing deactivation channels for Ir(III) complexes are illustrated in Scheme **1a**, whereas the corresponding kinetic model is presented in Scheme **1b**. Having in mind this kinetic model the photoluminescence efficiencies (1) and lifetimes (2) can be expressed as

$$\Phi_P(T) = \frac{k_r}{k_r+k_{ISC}+k_{nr}(T)} \quad , \quad (1)$$

$$\tau(T) = \frac{1}{k_r+k_{ISC}+k_{nr}(T)} \quad , \quad (2)$$

where $k_r$ is the $^3$MLCT→S$_0$ radiative decay rate, $k_{ISC}$ is the $^3$MLCT→S$_0$ nonradiative intersystem crossing rate (ISC) which in principle is temperature dependent through vibration population but here it is much less pronounced because the relevant vibrational modes are of high frequency, and $k_{nr}(T)$ is the strongly temperature-dependent nonradiative decay rate associated with the population of the $^3$MC well. The recent progresses in theoretical methods and algorithms enabled the computation of phosphorescence decay rates at the first-principles level.[13,14,15] Despite these progresses, the calculations of all the nonradiative counterparts present in Eq. (1) has remained more cumbersome, especially the temperature-dependent decay rate, i.e., $k_{nr}(T)$. Having in mind that OLEDs should work at ambient temperatures, the accurate calculation of all decay rates is mandatory for practical applications. In fact, the two groups involved in the present work have provided preliminary kinetic models for estimating the photoluminescence efficiencies of red-to-green[16,17] and green-to-blue[18] Ir(III) complexes, respectively. A general approach valid for all kind of complexes regardless of their emission color, is still missing in the literature. Here, we fill this gap by reporting a general computational protocol to predict the photoluminescence lifetimes and efficiencies of Ir(III) complexes at any given temperature. In contrast to previous models, this protocol is in principle valid for all Ir(III) complexes, regardless of their emission color, and it is potentially transferable to other phosphor materials with similar excited state features, e.g., square planar Pt(II) complexes. Importantly, this protocol does not rely on simplified kinetic models and does not require knowing beforehand any experimental information about the complexes, so that it can be integrated in the automated *in silico* prescreening of promising PhOLEDs materials.

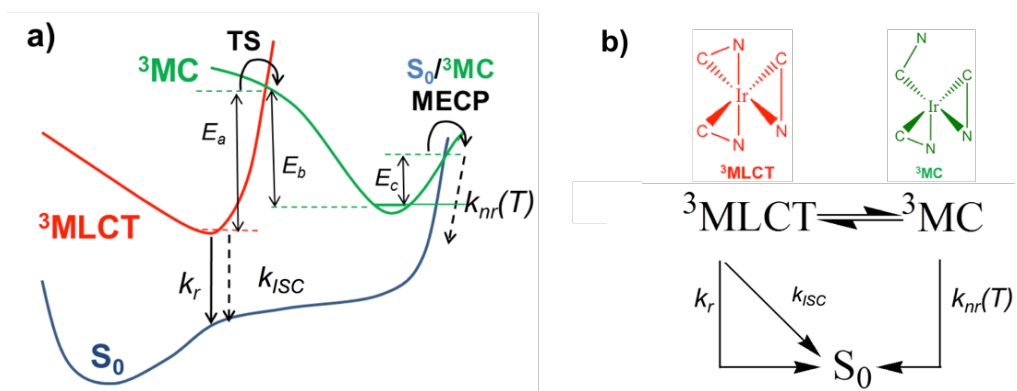

**Scheme 1. a)** Schematic representation of the competing excited state deactivation processes of Ir(III) complexes: emission from the $^3$MLCT state ($k_r$) competes with the $^3$MLCT→S$_0$ nonradiative decay ($k_{ISC}$) and with thermally activated nonradiative decay ($k_{nr}$(T)). The latter channel is characterized by population of the $^3$MC well through a transition state (TS) and the irreversible recovery of the ground state (S$_0$) geometry through the S$_0$/$^3$MC minimum energy crossing point (MECP). Along this channel, the main geometrical changes occur from the $^3$MLCT (pseudo-octahedral) to the $^3$MC (trigonal bipyramid) geometry, through breaking of one Ir-N bond (see Scheme 1b, top). Small geometrical changes occur along the $^3$MC→MECP transformation, the latter geometry exhibiting a further distorted trigonal bipyramid arrangement. Thus, the barrier to populate the TS ($E_a$) is usually the rate limiting step of this channel. For more details see Ref. 18. **b)** Kinetic model for the excited-state deactivation of Ir(III) complexes. Note that singlet excited states are not included, as the intersystem crossing processes present a 100% efficiency in these complexes.[11]

To test the validity of this protocol we have selected three Ir(III) complexes (see Scheme **2**), i.e., *fac*-Ir(F$_2$ppy)(ppz)$_2$ (**1**, where ppy = 2-phenylpyridine and ppz = phenylpyrazole), *fac*-Ir(ppy)$_3$ (**2**) and *fac*-Ir(flpy)$_3$ (**3**, where flpy = 2-(9,9-dimethyl-9*H*-fluoren-2-yl)pyridine), that emit in the blue, green and yellow/orange regions, respectively; and for which extensive experimental data is available.[10,19,20] Herein, for **1**-**3**, the $k_r$ decay rates are determined with the Einstein relationship and the $k_{ISC}$ decay rates are computed using the thermal vibration correlation function (TVCF) rate theory in combination with time-dependent density functional theory (TD-DFT) and DFT calculations. Regarding the calculation of $k_r$, it is generally accepted that perturbative approaches yield systematically slightly lower rates than non-perturbative schemes[21] or those that include spin-vibronic countributions.[17] However, these two latter sophisticated approaches are computationally more demanding, and would limit the general applicability of our protocol. Moreover, it was previously found for Ir(III) complexes,[22] that an optimal computational protocol based on a perturbative scheme can still provide $k_r$ values in good agreement with the experimental ones and is therefore selected herein (see details in the Computational Methods). The computation of the $k_{ISC}$ decay rates using the TVCF formalism has previously shown to yield robust results.[17,23] Finally, the $k_{nr}(T)$ decay rate is computed with canonical variational transition state theory (CVT),[24] based on the computed DFT energy profiles for the thermally activated nonradiative decay (see Scheme **1a**). As recently reported for these complexes,[18] the most suited kinetic model for this decay channel involves equilibration between the $^3$MLCT and $^3$MC states before irreversible return to the ground state. Thus,

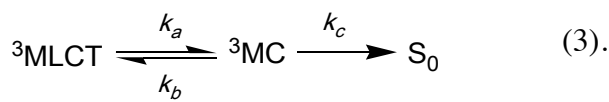

$$^3\text{MLCT} \underset{k_b}{\overset{k_a}{\rightleftharpoons}} {}^3\text{MC} \xrightarrow{k_c} S_0 \qquad (3).$$

Using the steady state approximation and assuming this complex kinetic scenario, the $k_{nr}(T)$ decay rate has the following form

$$k_{nr}(T) = k_c k_a/(k_c + k_b) \quad (4),$$

where $k_a$, $k_b$ and $k_c$ (i.e., $k_n$) can be expressed as

$$k_n = A_n \exp(-E_n/k_B T) \quad (5),$$

$E_n$ being the activation energy, $A_n$ its corresponding preexponential factor, and $k_B$ the Boltzmann constant. Computing all rates in Eq. (4) becomes easily prohibitive for practical purposes. As detailed in Ref. 18, the rate limiting step is usually the $^3$MLCT→$^3$MC transformation (i.e., $k_a$), which involves the breaking of one Ir-N bond (see Scheme **1b**). Under these circumstances, Eq. (4) can straightforwardly be approximated as

$$k_{nr}(T) = A_0 A \exp(-E_a/k_B T) \quad (6),$$

where $A_0$ stands for a temperature-dependent Boltzmann prefactor in accordance to (4), i.e.,

$$A_0 = \frac{1}{1+\exp\left(\frac{E_c-E_b}{k_B T}\right)} \quad (7),$$

and $A$ is the preexponential factor for the MLCT→$^3$MC transformation, which is computed with the CVT dynamics. Eq. 6 is thus capable to cover both a pre-equilibrated scenario as well as a non pre-equilibrated one. Complete details on the TVCF and CVT procedures can be found in the Methods section. Below we first discuss the blue emitter (**1**) in details, before presenting the results for the green (**2**) and yellow/orange (**3**) emitters.

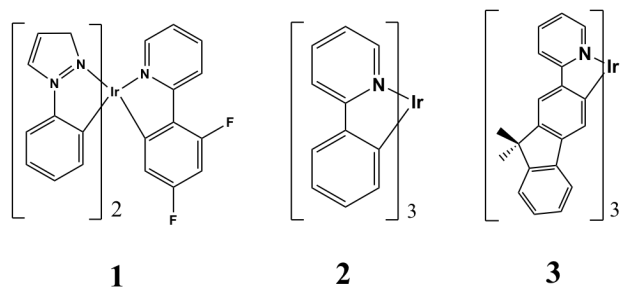

**Scheme 2.** Chemical structure of complexes **1**-**3**.

**Results and discussion**

**Blue-emitter case [*fac*-Ir(F$_2$ppy)(ppz)$_2$ (1)]:** The temperature-dependent photoluminescence properties of complex **1** were exhaustively investigated by Thompson and coworkers.[10] Relevant experimental photoluminescence properties of **1** at cryogenic and room temperatures (RT) are collected in Table 1. Experimentally, this blue emitter undergoes a significant drop of efficiency and lifetime at 298 K. As corroborated both experimentally[10] and theoretically,[18] this is due to the population of the thermally activated nonradiative decay process, which is readily accessible for **1** at RT as a result of its reduced $E_a$ value. Figure **1a** (**1b**) displays the recorded (computed) plot of the global decay lifetime versus temperature and the computed decay components at different temperatures are presented in Table 2. An overall qualitative and quantitative agreement of the experimental decay lifetime is recovered by the calculations. Importantly, the sigmoid-like dependence is recovered by the computed profile. In the following we will study in detail the origins of the lifetime's drop with increasing temperatures. In these plots, two distinct regimes are observed, a slight lifetime decrease at low temperatures and a second more significant drop at RT due to activation of the $k_{nr}(T)$ decay channel. Hence, two important inflexion points are observable, which are computed at ca. 220 K and 320 K, and experimentally observed at 250 K and 340 K. A deeper analysis of the results (see Table 2) reveals that before the first inflexion point, in the low-temperature regime (100-200 K), only the $k_r$ and $k_{ISC}$ components contribute to the global lifetime (see Table 2). Due to the large $^3$MLCT→S$_0$ SOCs, $k_r$ are of the order of $10^5$. Regarding the $k_{ISC}$ values, they are notably influenced by the $^3$MLCT-S$_0$ energy gap and by the total reorganization energies. The total reorganization energy, which is a measure of the extent of vibronic coupling between these two electronic states, it is obtained as a sum over all normal mode energies times their corresponding Huang-Rhys factor. Figure 2 collects the reorganization energies of the most relevant normal modes. The normal modes in the 500-1600 cm$^{-1}$ region predominantly contribute to the $k_{ISC}$ value of **1** (see the displacement vectors of its three most relevant modes in Figure 2, which involve aromatic C=C stretchings and in-plane deformations of mainly the (F$_2$ppy) ligand). As the relevant vibrational modes are of high frequency, there is only a slight increase of $k_{ISC}$ with temperature (see Table 1). Finally, $k_{nr}(T)$ originates from the thermally-activated formation of the $^3$MC state through the rupture of one Ir-N coordinating bond (see discussion above). In order to perform the CVT dynamics, the relevant stationary points of the rate limiting step, i.e., $^3$MLCT→TS→$^3$MC, were optimized and subsequently intrinsic reaction coordinate (IRC) calculations were carried out in both forward and backward directions (B3LYP/6-31G(d,p)/LANL2DZ). In Figure 3 the corresponding IRC including selected geometries along the path are plotted. The $^3$MLCT→TS→$^3$MC path is characterized by two main reaction coordinates: Ir-N bond stretching (from 2.18 to 3.43 Å) and an increase in the $\phi_{1,2,3,4}$ dihedral angle (from 3 to 60° see Figure 3). Herein, one critical aspect that needs to be mastered for the CVT simulations is the accurate estimation of $E_a$; since, due to their exponential relationship, small deviations in the energies imply huge changes in the decay rates (see Eq. (6)). Towards this end, and based on previous studies on the activation energies of 5d transition metal complexes,[25] the double-hybrid functional including dispersion PWPB95-D3 is chosen herein.

The results of the CVT simulations are collected in Table 3. For **1**, the computed activation barrier of the limiting step ($^3$MLCT→$^3$MC) is 2836 cm$^{-1}$. Experimentally, a Boltzmann analysis of the temperature-dependent photoluminescent decay data was performed and the activation barrier was determined to be 3300 cm$^{-1}$.[10] To illustrate the problematic in the accuracy needed for the rate calculations, the activation barrier was determined to be ca. 1097 cm$^{-1}$ with the B3LYP functional,[18] clearly underestimating the PWPB95-D3 value and the experimental evidences. The exponential prefactor $A$ obtained with the CVT simulations is $1.47 \times 10^{13}$ s$^{-1}$. The $E_{b-c}$ values as well as the Boltzmann $A_0$ prefactors at relevant temperatures are listed in Table S1. The $k_{nr}(T)$ values computed at different temperatures with Eq. (6) are collected in Table 2. Coming back to the temperature-dependent photoluminescence decay of **1**, the slight lifetime decrease in the 100-200 K regime can be understood as resulting from i) the increasing thermal population of the triplet substate with the largest $k_r$ value among the three substates of $^3$MLCT (see the three spin sublevel contributions in Table S2) and ii) the slight increase of $k_{ISC}$ with increasing temperatures. In this temperature regime, the computed photoluminescence efficiency achieves a maximum value at 196 K (i.e., 0.38, see Table 2). Importantly, the $k_{nr}(T)$ decay channel becomes only accessible at temperatures above ca. 230 K (see Table 2 and Figure 1). At RT it becomes the main nonradiative channel and directly competes with radiative decay. Finally, with increasing temperatures (up to 400 K) $k_{nr}(T)$ evolves into the most prominent deactivation channel (see Table 2) and it is thus responsible for the total quench of the photoluminescence at high temperatures. The relevance of the $k_{nr}(T)$ channel is further illustrated in Figure **1b**. Hence, by switching off the contribution of the $k_{nr}(T)$ component to Eq. (2) (dashed line in Figure **1b**), the sigmoid-like temperature-dependence is lost. Regarding the efficiencies, our calculations predict a photoluminescence efficiency of 0.38 and 0.05 at low temperatures and RT temperatures; respectively, in agreement with the experimentally observed drop in efficiencies. The theoretical $\Phi_P$ value at RT of **1** is underestimated with respect to the experimental one, likely due to the computational inaccuracies on the rate calculations (see exemplarily the discussion above regarding the energetic inaccuracies and their exponential relationship in the $k_{nr}(T)$ calculations).

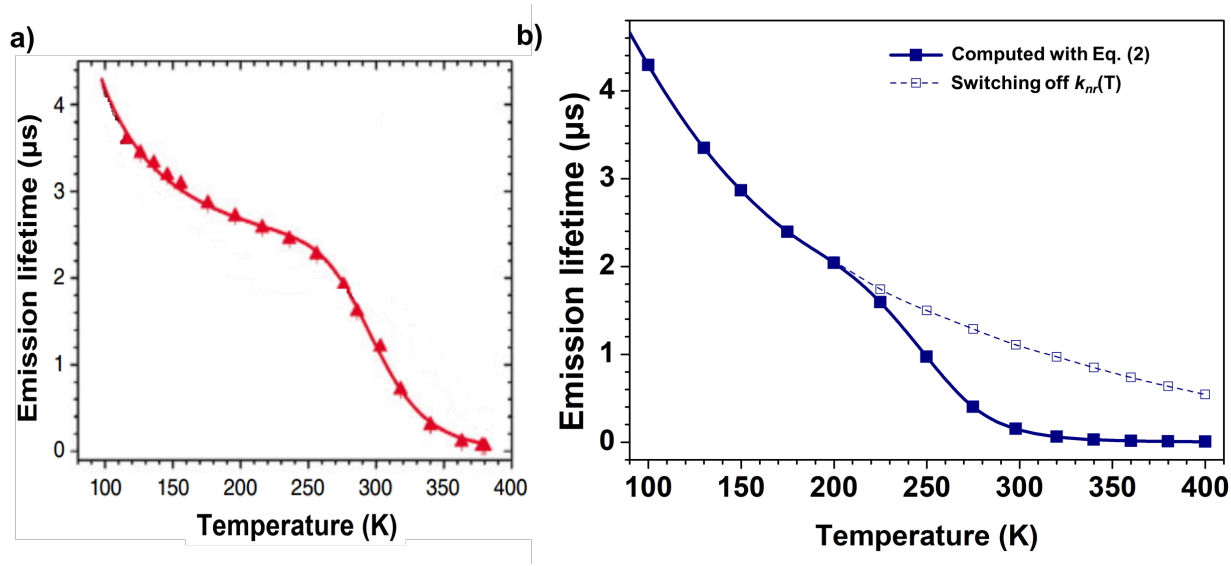

Figure **1**. Experimental (**a**) and computed (**b**) temperature dependence of the photoluminescence decay using Eq. (2) (bold line) and switching off the $k_{nr}(T)$ component (dashed line) for **1**. Figure 1a is adapted with permission from ref. 10. Copyright 2017 American Chemical Society.

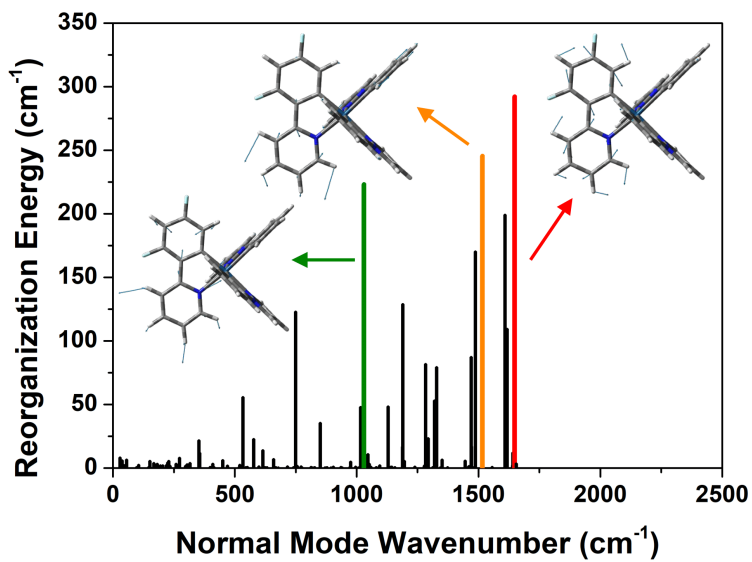

Figure 2. Calculated reorganization energies of each normal mode for 1. The displacement vectors of its three most relevant modes are depicted.

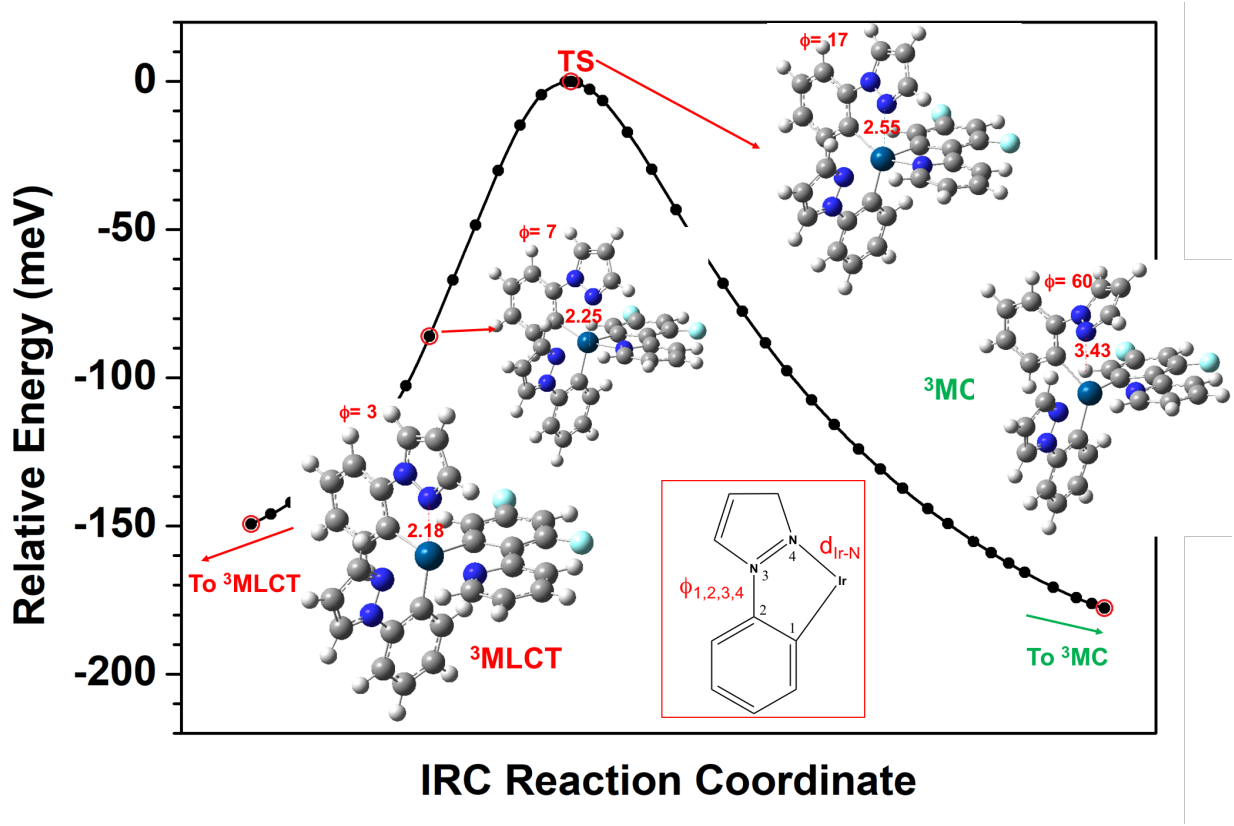

IRC Reaction Coordinate

Figure 3. Computed IRC for the $^3$MLCT→TS→$^3$MC pathway of **1** (B3LYP/6-31G(d,p)/LANL2DZ). C, N, H, F atoms in grey, dark blue, white and light blue, respectively.

Table 1. Experimental and theoretical photoluminescence properties of 1-3.

| Complex | $k_r$ (exp, s$^{-1}$) $k_{nr}$ (exp, s$^{-1}$) | $k_r$ (theo, s$^{-1}$) $k_{nr}$ (theo, s$^{-1}$)$^d$ | Room Temperature $\tau$ (exp, μs) $\tau$ (theo, μs) | $\Phi_P$ (exp) $\Phi_P$ (theo) | $k_r$ (theo, s$^{-1}$) $k_{nr}$ (theo, s$^{-1}$)$^d$ | Cryogenic Temperature$^c$ $\tau$ (exp, μs) $\tau$ (theo, μs) | $\Phi_P$ (exp) $\Phi_P$ (theo) |
|---|---|---|---|---|---|---|---|
| **1**$^a$ | 4.6×10$^5$ (3.8×10$^5$) | 3.0×10$^5$ (6.3×10$^6$) | 1.2 0.2 | 0.55 (0.05) | 1.1×10$^5$ (2.4×10$^5$) | 3.4 2.9 | -$^e$ (0.38)$^c$ |
| **2**$^a$ | 6.1×10$^5$ (1.9×10$^4$) | 4.1×10$^5$ (1.6×10$^5$) | 1.6 1.7 | 0.97 (0.72) | 1.9×10$^5$ (5.5×10$^4$) | 4.0 4.0 | -$^e$ (0.80)$^c$ |
| **3**$^b$ | 2.5×10$^5$ (6.0×10$^5$) | 2.5×10$^5$ (2.8×10$^6$) | 1.2 0.3 | 0.29 (0.08) | 2.3×10$^5$ (4.6×10$^5$) | 4.0 1.4 | -$^e$ (0.24)$^c$ |

$^a$Experimental values from ref. 10 in 2-MeTHF. $^b$Experimental values from ref. 20 in toluene at RT and on toluene/ethanol/methanol (5:4:1) solutions at cryogenic temperatures. $^c$Theoretical values for lifetimes computed at 150K and for efficiencies computed at 196 K. $^d$Obtained as the sum of all possible nonradiative decays. $^e$Not determined experimentally.

Table 2. Computed $k_r$, $k_{ISC}$ and $k_{nr}(T)$ decay rates, global lifetimes and $\Phi_P$ values at different temperatures for **1**.

| Temperature (K) | $k_r$ (s$^{-1}$) | $k_{ISC}$ (s$^{-1}$) | $k_{nr}(T)$ (s$^{-1}$) | Global Lifetime (μs) | $\Phi_P$ |
|---|---|---|---|---|---|
| 77 | 1.5994×10$^4$ | 1.7781×10$^5$ | 4.433E-12 | 5.1598 | 0.083 |
| 100 | 3.9948×10$^4$ | 1.9308×10$^5$ | 1.922E-06 | 4.2914 | 0.171 |
| 130 | 8.2342×10$^4$ | 2.1636×10$^5$ | 4.307E-02 | 3.3478 | 0.276 |
| 150 | 1.1326×10$^5$ | 2.3581×10$^5$ | 3.667E+00 | 2.8648 | 0.324 |
| 175 | 1.5163×10$^5$ | 2.6581×10$^5$ | 2.255E+02 | 2.3942 | 0.363 |
| 196 | 1.8236×10$^5$ | 2.9713×10$^5$ | 4.933E+03 | 2.0405 | 0.384 |
| 225 | $2.2156\times10^5$ | $3.5286\times10^5$ | 5.366E+04 | 1.5921 | 0.353 |
| 250 | $2.5211\times10^5$ | $4.1557\times10^5$ | 3.616E+05 | 0.9716 | 0.245 |
| 275 | $2.7976\times10^5$ | $4.9626\times10^5$ | 1.716E+06 | 0.4013 | 0.112 |
| 298 | $3.0281\times10^5$ | $6.0007\times10^5$ | 5.691E+06 | 0.1516 | 0.046 |
| 320 | $3.2292\times10^5$ | $7.0711\times10^5$ | 1.522E+07 | 0.0615 | 0.020 |
| 340 | $3.3971\times10^5$ | $8.3831\times10^5$ | 3.329E+07 | 0.0290 | 0.010 |
| 360 | $3.5522\times10^5$ | $1.0011\times10^6$ | 6.667E+07 | 0.0147 | 0.005 |
| 380 | $3.6957\times10^5$ | $1.2036\times10^6$ | 1.240E+08 | 0.0080 | 0.003 |
| 400 | $3.8287\times10^5$ | $1.4561\times10^6$ | 2.166E+08 | 0.0046 | 0.002 |

**Table 3. PWPB95-D3 activation barriers (cm$^{-1}$) for the rate-limiting step of the temperature-dependent non-radiative channels and $A$ exponential prefactor (s$^{-1}$) for 1-3.**

| Complex | $E_a$ (cm$^{-1}$) | $A$ (s$^{-1}$) |
|---|---|---|
| **1** | 2836[a] | $1.47\times10^{13}$ |
| **2** | 3676 | $1.70\times10^{13}$ |
| **3** | 5182 | $1.97\times10^{13}$ |

[a] The rate limiting step for **1** corresponds to the TS connecting $^3$MC→$^3$MC'. For **1** a more complex thermally activated deactivation channel is found involving $^3$MLCT, $^3$MC, $^3$MC' and MECP.

**Green-emitter case [*fac*-Ir(ppy)$_3$ (2)]:** Complex **2** has been extensively studied from both experimental[10,19] and computational viewpoints.[15,18,21,22,26] Therefore, it is the perfect scenario to benchmark our computational approach and to assess its temperature-dependent photoluminescent properties, which have rarely been the target of computational investigations. In Table S3 are collected the computed $k_r$, $k_{ISC}$ and $k_{nr}$(T) decay rates at different temperatures for **2**, whereas the computed plot of the global decay lifetime *versus* temperature can be found in Figure **4**. In this plot two inflexion points are observable, at ca. 280 K and 370 K. In the first temperature regime (100-280 K) there is only direct competition between the $k_r$ and $k_{ISC}$ components, since the computed $k_{nr}$(T) values are negligible (see Table S3). The $k_r$ values are of the same order than those reported in previous works.[15,22] The $k_{ISC}$ decay is mainly promoted by C=C stretching vibration modes and in-plane bending deformations, which are predominantly localized on a single ppy ligand[17] (see in Fig S1 the reorganization energies of the most relevant normal modes for **2**). The IRC of its rate-limiting step, i.e., $^3$MLCT→TS→$^3$MC, is shown in Figure S2. The main geometrical changes along this path are analogous to those of **1**. The computed $E_a$ value is 3676 cm$^{-1}$ (see Table 3) and the prefactor $A$ determined by the CVT dynamics is 1.7×10$^{13}$ s$^{-1}$. The $E_{b\text{-}c}$ values and the Boltzmann $A_0$ factors at relevant temperatures are listed in Table S1. Note that the computed barrier for **2** is, in agreement with the experiments, larger than that of **1**. The $k_{nr}$(T) component starts to be non-negligible only beyond RT (compare the dash and bold computed profiles in Figure **4**), and therefore, experimentally, **2** is still highly efficient at RT ($\Phi_P$=0.97). This is also corroborated by our calculations, which yield a value of 0.72 at RT. This value is quite close to the maximum $\Phi_P$ value in the low-temperature regime i.e., 0.80 at 196 K (see Table 1). Thus, the activation barrier to populate the $^3$MC channel is large enough to prevent its population in the 100-280 K regime. Conversely, the $A$ value is of the same order for both complexes. Thus, the larger $E_a$ value for **2** is the main feature that prevents the population of the $k_{nr}$(T) decay channel up to ambient temperatures and hence ensures its large $\Phi_P$ value. Above 300 K, the radiative efficiency of complex **2** decreases significantly (see Table S3).

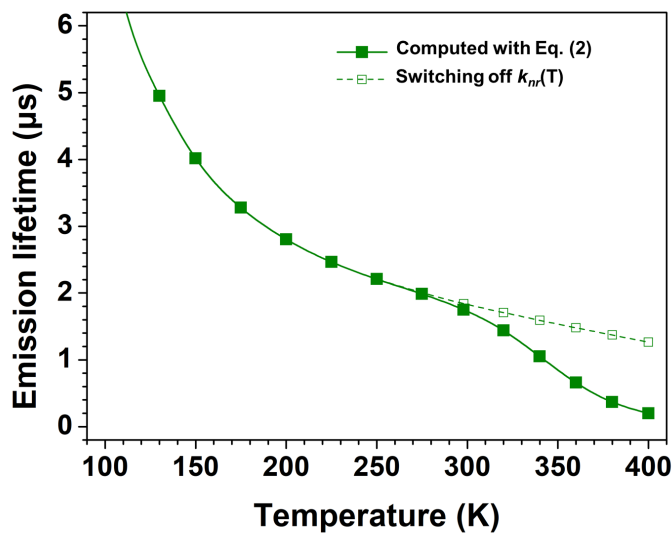

Figure **4.** Computed temperature dependence of the photoluminescence decay using Eq. (2) (bold line) and switching off the $k_{nr}(T)$ component (dashed line) for **2**.

**Yellow/Orange-emitter case [*fac*-Ir(flpy)$_3$(3)]:** Due to their reduced T$_1$-S$_0$ energy gap; yellow, orange and red Ir(III) complexes are, in this increasing order, more prone to T$_1$→S$_0$ nonradiative decay than green and blue ones. Conversely, this trend is reversed for the $k_{nr}(T)$ component, since a more energetically stabilized T$_1$ state generally leads to a larger $E_a$ value that prevent the population of the $^3$MC state for the former complexes. The yellow/orange complex **3** was studied in detail from an experimental point of view by Tsuboyama and coworkers.[20] Relevant experimental and computed photoluminescence data of **3** at cryogenic and room temperatures (RT) are collected in Table 1. In Table S4 the computed $k_r$, $k_{ISC}$ and $k_{nr}(T)$ decay rates at different temperatures are listed, whilst the plot of the computed global decay lifetime *versus* temperature is found in Figure **5**. The IRC of its $^3$MLCT→TS→$^3$MC pathway is shown in Figure S4 and it shows identic geometrical distortions to those of **1-2**. Finally, the computed $E_a$ and $A$ values are collected in Table 3 whilst the $E_{b-c}$ and $A_0$ values are reported in Table S1.

Complex **3** possesses the largest $E_a$ value among all the studied complexes (5182 cm$^{-1}$), thus confirming the initial hypothesis of diminished relevance of the $k_{nr}(T)$ component for yellow-to-red complexes. Indeed, the computed $k_{nr}(T)$ values are negligible in all temperature regimes (77-400 K), see Table S4; and thus the effect of switching off this component has a negligible effect (see in Figure **5**). Therefore, the photoluminescence lifetimes of **3** are only determined by a direct competition between the $k_r$ and $k_{ISC}$ components. The computed $k_r$ value (2.5×10$^5$) shows a perfect agreement with its experimental counterpart (see Table 1). The $k_{ISC}$ is promoted by similar aromatic modes of those of **1-2**, but conversely mainly localized on two ligands (see reorganization energies and the most relevant modes in Figure S3). Note that this complex is characterized by the smallest total reorganization energies amongst all the complexes. The agreement between the experimental and computed $\Phi_P$ value is remarkable (see Table 1). Overall, and despite the negligible $k_{nr}(T)$ contribution, smaller $\Phi_P$ values are obtained for **3** with respect to **1-2**. This trend is observed both computationally and experimentally and it is due to the confluence of two factors, a smaller $k_r$ value and an enhanced $k_{ISC}$ value for **3**. The smaller $k_r$ value can be rationalized in terms of the more pronounced ligand-centered ($^3$LC) character of T$_1$, that leads to smaller SOCs and ZFS values (compare the ΔE splitting between triplet sublevels for all complexes in Table S2). The enhanced $^3$LC character for **3** is corroborated by the small contribution of the iridium atom to the spin density distribution at the T$_1$ (see Figure S7). Additionally, and despite the small total reorganization energies for **3**, its reduced T$_1$-S$_0$ energy gap leads to an enhanced $k_{ISC}$ decay rate as compared to **1-2**. Therefore, these observations are in accordance to the energy-gap law.

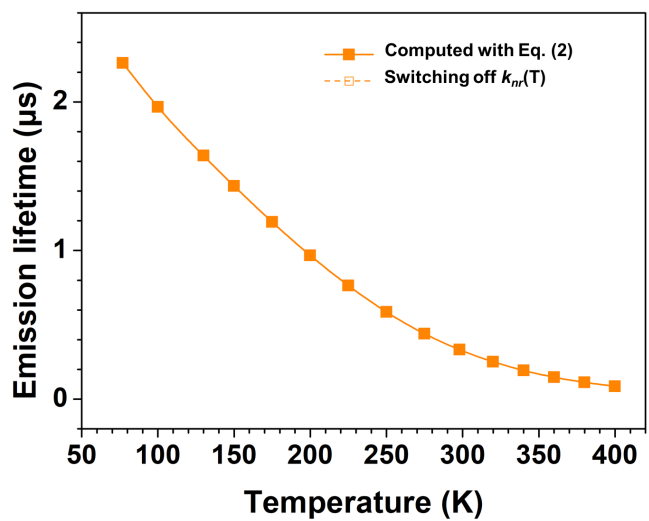

Figure **5.** Computed temperature dependence of the photoluminescence decay using Eq. (2) (bold line) and switching off the $k_{nr}(T)$ component (dashed line) for **3**. Due to the negligible contribution of the $k_{nr}(T)$ component, note that both computed profiles are coincident.

**Methods**

All the molecular parameters needed in our model are calculated by density functional theory (DFT). The geometries of the singlet ground state ($S_0$), the $^3$MC and $^3$MLCT triplet excited states, and the transition state (TS) were optimized for complexes **1**-**3** using the hybrid functional B3LYP, in combination with the 6 31G(d,p) atomic basis set for all first and second-row atoms. Relativistic effects were included for the Ir atom by using the LANL2DZ pseudopotential. The nature of the stationary points was confirmed by computing the Hessian at the same level of theory. In the case of complex **3**, the two methyl positions of the 9*H*-fluoren-2-yl)pyridine unit were replaced by hydrogen atoms for computational ease. The spin density distributions at the $^3$MLCT, $^3$MC and TS optimized geometries of **1**-**3** are depicted in Figures S3-5. The $S_0$/$^3$MC MECP was optimized using Harvey's algorithm, as implemented in the ORCA software;[27]

employing the B3LYP functional in combination with the *def2*-SVP atomic basis set and the ECP-60-mwb Stuttgart/Dresden pseudopotential for Ir. To get accurate energies of $E_a$, single point calculations were performed with the dispersion-corrected double-hybrid PWPB95-D3 functional,[28] in combination with the *def2*-SVP atomic basis and the ECP-60-mwb Stuttgart/Dresden pseudopotential for Ir, as implemented in ORCA. All calculations apart from the MECP optimization and the double-hybrid calculations, were carried out with the Gaussian09 program package.[29] The effect of solvation on the geometries and on the computed activation barriers was assessed using THF as a solvent and the polarizable continuum model (PCM)[30] as implemented in Gaussian. These results are presented in Table S5 for **1**. In view of the small differences observed, the gas phase results were used in all the calculations for all complexes.

To model the $k_{nr}(T)$ decay channel, CVT dynamics were performed. Towards this end, intrinsic reaction coordinate (IRC) calculations of the rate limiting step, i.e., $^3$MLCT→TS→$^3$MC, were carried out at the B3LYP/6-31G(d,p)/LANL2DZ level. Each point along the IRC path was used to build a reliable $^3$MLCT→TS→$^3$MC potential energy surface (PES). The CVT dynamics were done with the POLYRATE suite of programs.[31]

The spin-orbit coupling (SOC) was treated as a perturbation and the exact two-component (X2C) Hamiltonian was applied to construct the SOC operator. For the $k_r$ and $k_{ISC}$ calculations, the SOC calculations were carried out at optimized $S_0$ geometry with the BDF program package[32] using TD-B3LYP in combination with the 6-31G(d,p) basis set for light atoms and the ANO-RCC-VDZP basis set for iridium atom.

The radiative decay rates of spin sublevels ($k_{r,i}$) are evaluated using Einstein spontaneous emission formula

$$k_{\mathrm{r},i} = \frac{f_i E_{\mathrm{vert},i}^2}{1.499 \text{ s}\cdot\text{cm}^2} \qquad (7),$$

where subscript $i$ denotes the spin sublevel along with its oscillator strength ($f_i$) and vertical transition energy ($E_{\mathrm{vert},i}$), which are computed at the level of theory stated above using the BDF program package. Boltzmann statistics of these three spin sublevels are then applied to calculate the overall radiative decay rate ($k_r$) at different temperature

$$k_{\mathrm{r}} = \frac{k_{\mathrm{r,I}} + k_{\mathrm{r,II}}\exp(-E_{\mathrm{II,I}}/k_B T) + k_{\mathrm{r,III}}\exp(-E_{\mathrm{III,I}}/k_B T)}{1 + \exp(-E_{\mathrm{II,I}}/k_B T) + \exp(-E_{\mathrm{III,I}}/k_B T)} \qquad (8),$$

where $E_{\mathrm{II,I}} = E_{\mathrm{vert,II}} - E_{\mathrm{vert,I}}$ and $E_{\mathrm{III,I}} = E_{\mathrm{vert,III}} - E_{\mathrm{vert,I}}$ are the energy difference between spin sublevels. This thermal rearrangement between the triplet sublevels of $^3$MLCT originates the temperature-dependency of the $k_r$ values at cryogenic temperatures.

For Ir(III) complexes, which are characterized by large SOCs, first-order perturbation theory is a good approximation to compute the nonradiave ($k_{\mathrm{ISC}}$) rate. The $k_{\mathrm{ISC}}$ is evaluated using the TVCF theory. Perturbation theory was also applied to compute the non-adiabatic electronic coupling and the spin-orbit couplings. The TVCF formalism makes use of a multidimensional harmonic oscillator model coupled with DFT and TD-DFT calculations, where distortions, displacements and Duschinsky rotations are taken into account. Thus, the difference between two electronic state potential energy surfaces was considered by using $\underline{Q}_e = S\ \underline{Q}_g + \underline{D}_e$, where $S$ is the Duschinsky rotation matrix, $\underline{Q}_e$ and $\underline{Q}_g$ are the normal-mode coordinates vectors of the $S_0$ and $T_1$ states, respectively; and $\underline{D}_e$ is the displacement vector between the minima of the excited and ground state geometries. Further details can be found in Ref. 17. The three key parameters governing the $k_{ISC}$ values are: i) the adiabatic energy gap, ii) the total reorganization energy and iii) the size of SOCs. The $k_{ISC}$ rate calculations were performed with the MOMAP suite of programs was used.[16,33,34,35,36,37]

**Conclusions**

The calculation of the temperature-dependent photoluminescent properties of phosphors has remained up to date elusive. In this manuscript, we report for the first time a general approach to compute the temperature-dependent photoluminescence lifetimes and efficiencies of Ir(III) complexes. Since all possible kinetic scenarios are included within our approach, its validity is not restrained for a limited series of complexes (like previous schemes), but it remains valid for a wide chemical space, i.e., from deep-blue to yellow-to-red phosphors. Importantly, the photoluminescence lifetimes and efficiencies can be estimated at any given temperature and the temperature-dependent photoluminescence decay profiles can be derived. Our future work should be devoted to further improve the efficiencies' estimations in a quantitative basis, which it remains difficult due to the computational inaccuracies on the energetics and their exponential relationships in the rate calculations.

Herein, we demonstrate that the strongly-temperature dependent $k_{nr}(T)$ channel is non-negligible up to green complexes. Accordingly, this channel is increasingly less relevant for yellow, orange and red complexes. In the case of blue complexes, controlling the $k_{nr}(T)$ channel is mandatory to attain large photoluminescence efficiencies at RT. This can be accomplished *via* stronger Ir-heteroatom bonds that lead to larger $E_a$ values, such as e.g., in N-Heterocyclic Carbene (NHC)-based Ir(III) complexes. For future blue phosphor materials design, and in view of the computed evidences, an $E_a$ of ca. 4000 cm$^{-1}$ might be enough to impede the population of the $k_{nr}$(T) channel at RT. Additionally, this strategy is also beneficial to attain photostable blue Ir(III) emitters.[38] Due to the enormous industrial efforts in obtaining highly-efficient phosphors and the increased use of high-throughput approaches on computational materials' design, we anticipate that this protocol will be of great importance for the automatic *in silico* prescreening of promising PhOLEDs materials.

## ACKNOWLEDGMENTS

DE thanks funding from the European Union's Horizon 2020 research and innovation programme under the Marie Sklodowska-Curie grant agreement No 700961. DJ acknowledges the European Research Council (Marches grant n°278845) and the RFI Lumomat for financial support. The GRDI-RCTF network is acknowledged for funding short mission travel grants. The work in Beijing is supported by the Ministry of Science and Technology of China Through Grant No. 2017YFA0204501 and the National Natural Science Foundation of China Grant No. 91622121.

## SUPPORTING INFORMATION

Figures S1-S7, Tables S1-S22 including the xyz coordinates of all complexes are presented in the Supporting Information.

TOC Graphic

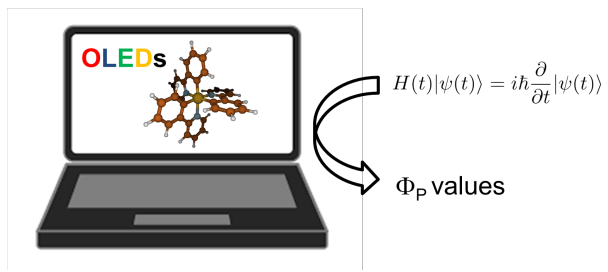

From first-principles computations of the photoluminescence lifetimes and efficiencies of Ir(III) complexes.